\documentclass[11pt]{article}
\usepackage{color}
\usepackage{epsfig}

\begin{document}

\begin{center}
{\Large\bf Holographic dark energy: constraints on the interaction from diverse observational data sets}
\\[15mm]
Purba Mukherjee \footnote{E-mail: pm14ip011@iiserkol.ac.in},
Ankan Mukherjee \footnote{E-mail: ankanju@iisermohali.ac.in},
H. K. Jassal \footnote{E-mail: hkjassal@iisermohali.ac.in},
Ananda Dasgupta \footnote{E-mail: adg@iiserkol.ac.in},
Narayan Banerjee \footnote{E-mail: narayan@iiserkol.ac.in}\\[5mm]

{\em $^{1,4,5}$ Department of Physical Sciences,~~\\Indian Institute of Science Education and Research Kolkata,\\ Mohanpur, West Bengal 741246, India.}\\[2mm]
{\em $^{2,3}$ Department of Physical Sciences,~~\\Indian Institute of Science Education and Research Mohali, \\ Sector 81, SAS Nagar, Mohali, Punjab 140306, India.}\\[15mm]
\end{center}

\vspace{0.5cm}
\vspace{0.5cm}
\pagestyle{myheadings}
\newcommand{\be}{\begin{equation}}
\newcommand{\ee}{\end{equation}}
\newcommand{\bea}{\begin{eqnarray}}
\newcommand{\eea}{\end{eqnarray}}

\begin{abstract}
The present work deals with holographic dark energy models with Hubble horizon as the infra-red cut-off. The interaction rate between dark energy and dark matter has been reconstructed with three different choices of the interaction term. It is shown that the coupling parameter of the interaction term should evolve with redshift to allow the successful transition from decelerated to accelerated phase of expansion. Constraints on the model parameters are obtained from Markov Chain Monte Carlo (MCMC) analysis using the supernova distance modulus data and observational measurements of the Hubble parameter. Results show that the model with the coupling parameter increasing with redshift ($z$) or equivalently decreasing with the evolution, are ruled out. On the other hand, coupling parameters, increasing or slowly varying with the evolution, are consistent with the observed evolution scenario. A Bayesian evidence calculation has been carried out for statistical selection of the reconstructed models. Though the kinematical parameters are well behaved for these models, the physical variables which determine the nature of the components in the matter sector, are not at all realistic. We have concluded that the existence of spatial curvature is essential for this particular type of dark energy models.
\end{abstract}

\vskip 1.0cm

PACS numbers: 98.80.Cq; 98.80.−k; 95.36.+x

Keywords: cosmology, dark energy, holographic principle.

\section{Introduction}
The exotic component, introduced in the energy budget of the universe to account for the phenomenon of cosmic acceleration is dubbed as {\it dark energy}. It is a theoretical prescription to accommodate the alleged accelerated expansion of the universe within the framework of General Relativity (GR). Cosmological observations are highly consistent with the {\it cosmological constant} model of dark energy where the vacuum energy density serves as the dark energy candidate. But it suffers from a fine-tuning problem. The observationally required value of the cosmological constant is very small compared to the value of vacuum energy density, estimated in quantum field theory, and their ratio is of the order of $10^{-120}$. Different aspects of cosmological constant models are discussed in great detail\cite{carrol, pady}. This huge discrepancy leads to the search for other forms of dark energy. The alternative options within the regime of GR are the time-varying dark energy models. These are also well consistent with cosmological observations. Time-varying dark energy can be described by scalar field models, like quintessence\cite{quint}, phantom field\cite{phantom}, tachyon field\cite{tachyon} or by fluid models like Chaplygin gas\cite{chap}.

\par In the present work, we emphasize another alternative description of dark energy, namely the holographic dark energy. The idea of holographic dark energy stems from thermodynamics, namely the {\it holographic principle}, discussed briefly in the following section. It is important to mention in this context that Horava and Minica\cite{harami} have argued that the holographic principle can potentially resolve the problem of fine-tuning of cosmological constant as they have shown that the holography theory implies that the most probable value of the cosmological constant is zero. The description of dark energy, inspired by the holographic principle, is characterized by a typical length scale of the system, called the infra-red (IR) cut-off. In the present context, it is actually of the cosmological horizon size, but the choice of the horizon is not quite unique.

\par In the context of dark energy, the holographic principle was first introduced by Li\cite{liholo} where the future event horizon has been utilized as the IR cut-off. There are several attempts in the literature with different IR cut-off, like particle horizon\cite{holopart}, future event horizon\cite{liholo,holofu}, Hubble scale IR cut-off\cite{holohub, xu}, Ricci scale cut-off\cite{holoricci} etc. A combination of the square of Hubble parameter and its time derivative was used to define the IR cut-off by Granda and Oliveros \cite{Granda:2008dk} and the same IR cut-off has been utilized further by Easson {\it et al} \cite{Easson:2010av} and by Basilakos and Sola \cite{Basilakos:2014tha}. 

\par A comparative study of holographic dark energy models with different length scale cut-off has been carried out by del Campo {\it et al.} \cite{delCampo}.  Hu {\it et al.} made an attempt to combine the cosmological constant with the holographic dark energy\cite{huholocos} (HDE). Evolution of matter perturbation in holographic dark energy models has been studied by del Campo {\it et al.}\cite{delCampo2}, Mehrabi {\it et al} \cite{mehrabi}, Malikjani {\it et al} \cite{Malekjani:2016edh}. Holographic dark energy in Brans-Dicke theory has been discussed by Banerjee and Pavon\cite{nbpavon}. Recently, Lamdin\cite{lamdin} has discussed HDE in the context of minimal super-gravity. Li {\it et al.}\cite{liplanck} has studied the constraints on holographic dark energy from recent Planck data. A stability analysis of holographic dark energy has been carried out by Banerjee and Roy \cite{nbnr} and by Mahata and Chakraborty\cite{mahatacha}. The phantom - non-phantom transition in connection with HDE was discussed by Nojiri and Odintsov\cite{odin1}. The same authors presented quite a general HDE model very recently \cite{odin2} which yields many HDE models as special cases. An attempt to unify the cold dark matter and dark energy fields through an interaction has been discussed by von Marttens {\it et al} \cite{vonMarttens:2018iav}.

\par In the present work, the Hubble horizon is considered as the IR cut-off of the holographic dark energy. In a spatially flat  Friedmann-Lemaitre-Robertson-Walker (FLRW) universe, holographic dark energy model with Hubble horizon cut-off essentially requires an interaction between dark energy and dark matter for a successful transition from decelerated to accelerated phase of expansion\cite{liholo, holohub}. A viable interaction between dark energy and dark matter and also to obtain the possible constraints on the interaction rate for this particular holographic dark energy scenario is looked for in this work.

\par Reconstruction of holographic dark energy interaction rate from parametrized dark energy equation of state has been discussed by Sen and Pavon \cite{senpav}. Interaction rate in holographic dark energy has been reconstructed through a kinematic approach by Mukherkjee \cite{am}. In both these cases, the interaction rate is reconstructed for some particular evolution scenarios. The present work is different as the evolutionary history is obtained from the parametrization of the interaction function itself. The interaction function is assumed to be proportional to $H\rho_H$, where $\rho_H$ is the holographic dark energy density and $H$ is the Hubble parameter. The proportionality parameter $\alpha$ determines the strength of the interaction. Praseetha and Mathew adopted this type of interaction in holographic dark energy to check the validity of the generalized second law of thermodynamics at the apparent and event horizon \cite{praseemathew} in interacting holographic models. Constraining the coupling parameter is the primary motivation of the present work. We have found that the coupling parameter characterizing the dark matter-dark energy interaction is not a constant in this case. It is essentially a function of time or redshift to allow the transition from decelerated to accelerated phase of expansion. We introduce a parameterized form of the coupling parameter. Three different parametrizations which evolve differently with redshift have been suggested. It will be seen that some parametrizations of the interaction do explain the observations, but none of them actually gives a clear picture of the constituents of the universe.

\par  This paper is arranged as follows. In section \ref{basicholo}, the basic holographic principle has been briefly reviewed. Section \ref{reconst} contains the reconstruction of the interaction term in holographic dark energy. The statistical analysis, including the discussion about the observational data, methodology and the results, are presented in section \ref{stat}. In section \ref{cospar}, the evolution of different cosmological parameters has been studied for the reconstructed models. In section \ref{bayev}, a Bayesian analysis for model selection has been carried out so as to pick up the best-suited parametrization. The results have been summarized with an overall discussion in section \ref{dis}.

\section{Basic holographic principle}
\label{basicholo}

't Hooft\cite{tHoo} and Susskind\cite{susholo} conjectured that the phenomena within a volume can be explained by the set of degrees of freedom residing on its boundary and the degrees of freedom are determined by the area of the boundary rather than the volume. This idea is based on the black hole entropy bound, suggested by Bekenstein \cite{beke}. The formation of a black hole leads to a connection between the short distance cut-off, namely the ultraviolet (UV) cut-off, to a long distance or IR cut-off\cite{cohen} by the constraint that the total quantum zero-point energy of the system should not exceed the mass of black holes of the same size. This can be expressed by the inequality as, 

\be 
L^3\rho_{\Lambda}\leq LM_p^2,
\ee 
where $M_p^2=(8\pi G)^{-1}$, $\rho_{\Lambda}$ is the quantum zero-point energy density determined by the UV cut-off and $L$ is the length scale of the size of the system. The length for which the inequality saturates is the long distance cut-off or the IR cut-off. In the context of dark energy, the holographic energy density is written as,

\be 
\rho_{H}=3C^2M_p^2/L^2,
\ee
where $C^2$ is a dimensionless coupling parameter \cite{liholo}.
For holographic dark energy, the system size is the observable universe and thus the IR cut-off is the cosmological horizon. The choice of the IR cut off is not unique. The reconstruction in the present work is carried out assuming the Hubble horizon as the IR cut-off of the holographic dark energy.

\section{Reconstruction of the interaction term}
\label{reconst}

The infinitesimal distance element in a homogeneous and isotropic universe is given by the Friedmann-Lema\^{i}tre-Robertson-Walker (FLRW) metric,
\be
ds^2=-dt^2+a^2(t)\left[\frac{dr^2}{1-kr^2}+r^2 d\theta^2 + r^2 \sin^2 \theta d\phi^2 \right],
\ee 
where, $a(t)$ is called the {\it scale factor} and $k$ is the curvature parameter. The present analysis has been carried out with the assumption of spatial flatness of the universe which implies $k=0$. The Hubble parameter is defined as, $H=\dot{a}/a$. The inverse of the Hubble parameter, which has a dimension of {\it time} (or equivalently length in natural units where $c=1$), represents the length scale, called the {\it Hubble horizon}. The Friedmann equations, in terms of Hubble parameter are given as,

\be
3H^2=8\pi G\rho_{tot},
\ee

\be
2\dot{H}+3H^2=-8\pi Gp_{tot},
\ee
where $\rho_{tot}$ and $p_{tot}$ are the energy density and pressure contributions of all the components in the energy budget of the universe respectively.

In the present work, the Hubble horizon has been taken as the IR cut-off length scale for the holographic dark energy, i.e. $L=\frac{1}{H}$.
Thus, the dark energy density is expressed as,

\be
\rho_{DE}=3C^2M_P^2H^2,
\ee
where $C^2$ is a dimensionless coupling parameter. The conservation equation of the total energy budget, obtained from the contracted Bianchi identity, is

\be
\dot{\rho}_{tot}+3H(\rho_{tot}+p_{tot})=0.
\label{consv}
\ee
At present, the prime contribution to the energy sector of the universe is coming from dark energy and pressure-less dark matter. Thus $\rho_{tot}$ can be written as, $\rho_{tot} = \rho_m + \rho_{DE}$. Finally the conservation equation (equation (\ref{consv})) can be separated into two parts, 
\be
\dot{\rho}_m+3H\rho_m=Q,
\label{matcon}
\ee 
and
\be
\dot{\rho}_{DE}+3H(1+w_{DE})\rho_{DE}=-Q.
\label{decon}
\ee
This $Q$ is the interaction function and $w_{DE}$ is the dark energy equation of state parameter defined as $w_{DE}=p_{DE}/\rho_{DE}$. For $Q=0$, these two equations (equation (\ref{matcon}) and (\ref{decon})) become decoupled allowing the independent conservation of dark energy and dark matter. In the present work, the interaction function $Q$ has been reconstructed with three different parametrizations. The general form of $Q$ is assumed to be $Q=3H\alpha(z)\rho_{DE}$, where the coupling term $\alpha $ is a function of redshift $z$. Now, let us define another quantity, which is called the {\it coincidence parameter (r)}, as $r=\rho_{m}/\rho_{DE}$. In case of holographic dark energy with Hubble horizon as the IR cut-off in a spatially flat universe, the coincidence parameter $r$ is a constant \cite{senpav}. For a constant $\alpha$, it can be shown from equation (\ref{decon}) that Hubble parameter $H\propto(1+z)^{\frac{3}{2}(1-\frac{\alpha}{r})}$. Hence the model cannot allow the transition from decelerated to accelerated phase of expansion. A time-varying coupling parameter $\alpha(z)$ is required for the successful transition from decelerated to accelerated phase of expansion. Here, three different ansatzes have been chosen for $\alpha(z)$ to reconstruct the interaction function $Q$, given as

\be
Model~~I.~~~~~\alpha(z)=\alpha_1+\alpha_2(1+z),
\label{a1}
\ee   
\be
Model~~II.~~~~~\alpha(z)=\alpha_1+\alpha_2\frac{z}{(1+z)},
\label{a2}
\ee  
\be
Model~~III.~~~~~~\alpha(z)=\alpha_1+\frac{\alpha_2}{(1+z)}
\label{a3}
\ee 

where $\alpha_1 , \alpha_2$ are constant parameters. It is customary to have some ansatz for the physical quantities in the reconstruction of models, and the parameters are estimated from the data sets. The rationale for the ansatz adopted here is to have simple but different modes of variation of the interaction with the evolution. Model I has a linear dependence on $z$, II has a mixed dependence and III has an inverse dependence.

The expressions of Hubble parameter obtained for these models are,

\be
I.~~~~~~~~~~H^2(z)=H^2_0\left[(1+z)^{3(1-\frac{\alpha_1}{r})}\exp{\left(-3\frac{\alpha_2}{r}z\right)}\right],
\label{hub1}
\ee   
\be
II.~~~~~H^2(z)=H^2_0\left[(1+z)^{3(1-\frac{\alpha_1+\alpha_2}{r})}\exp{\left(3\frac{\alpha_2}{r}\frac{z}{1+z}\right)}\right],
\label{hub2}
\ee  
\be
III.~~~~~~H^2(z)=H^2_0\left[(1+z)^{3(1-\frac{\alpha_1}{r})}\exp{\left(-3\frac{\alpha_2}{r}\frac{z}{1+z}\right)}\right],
\label{hub3}
\ee 

where $H_0$ is the present value of the Hubble parameter. It is important to mention in this context that the dark energy and dark matter terms cannot be separately identified in the equations for the Hubble parameter (equation (\ref{hub1}) to (\ref{hub3})) and the models cannot be reduced to non-interacting models. If $\alpha_1$ and $\alpha_2$ are set equal to zero, all the models reduce to a pure CDM model. As $r$ is constant for these models, we can redefine the parameters as $\beta_1=\alpha_1/r$ and $\beta_2=\alpha_2/r$. We can scale the Hubble constant ($H_0$) by $100~km.sec.^{-1}Mpc^{-1}$ to represent it in a dimensionless way as $h_0$. The parameters, which are constrained in present analysis, are ($h_0$, $\beta_1$, $\beta_2$). As the interaction function, $Q$ is now characterized by the parameters $\alpha_1$ and $\alpha_2$, the direction of energy flow in the interaction between dark energy and dark matter will be determined by the signature of the parameters $\alpha_1$ and $\alpha_2$. A positive $Q$ indicates the energy flow from dark energy to dark matter and a negative $Q$ indicates the reverse.

\section{Statistical analysis and constraints on the parameter}
\label{stat}

In the present analysis, two different data sets, namely the supernova distance modulus data and observational measurements of the Hubble parameter (OHD) have been utilized to constrain the model parameters.

\par The distance modulus measurements of type Ia supernova from the Joint Light-curve Analysis (JLA) \cite{jla} have been used in the present analysis. The observational measurements of Hubble parameter (OHD) at different redshift in the range $0.07<z<2.36$ by different groups have been taken into account.  The OHD, which are used in the present analysis, is normally measured by three different methods, Cosmic Chronometer method \cite{ohdcc}, measurements from galaxy distribution \cite{ohdbao} and from Lymann-$\alpha$ forest distribution \cite{ohdLya}.

\par The uncertainty of the parameters are estimated by the Markov Chain Monte Carlo (MCMC) method with the assumption of a uniform prior distribution. In Bayesian inference, the posterior probability distribution is proportional to the likelihood distribution of the parameter in case of a uniform prior. In the present analysis, we have adopted the \texttt{python} implementation of the ensemble sampler for MCMC, the \texttt{emcee}, introduced by Foreman-Mackey {\it et al.} \cite{emcee}.

\begin{figure}[tb]
\begin{center}
\includegraphics[angle=0, width=0.7\textwidth]{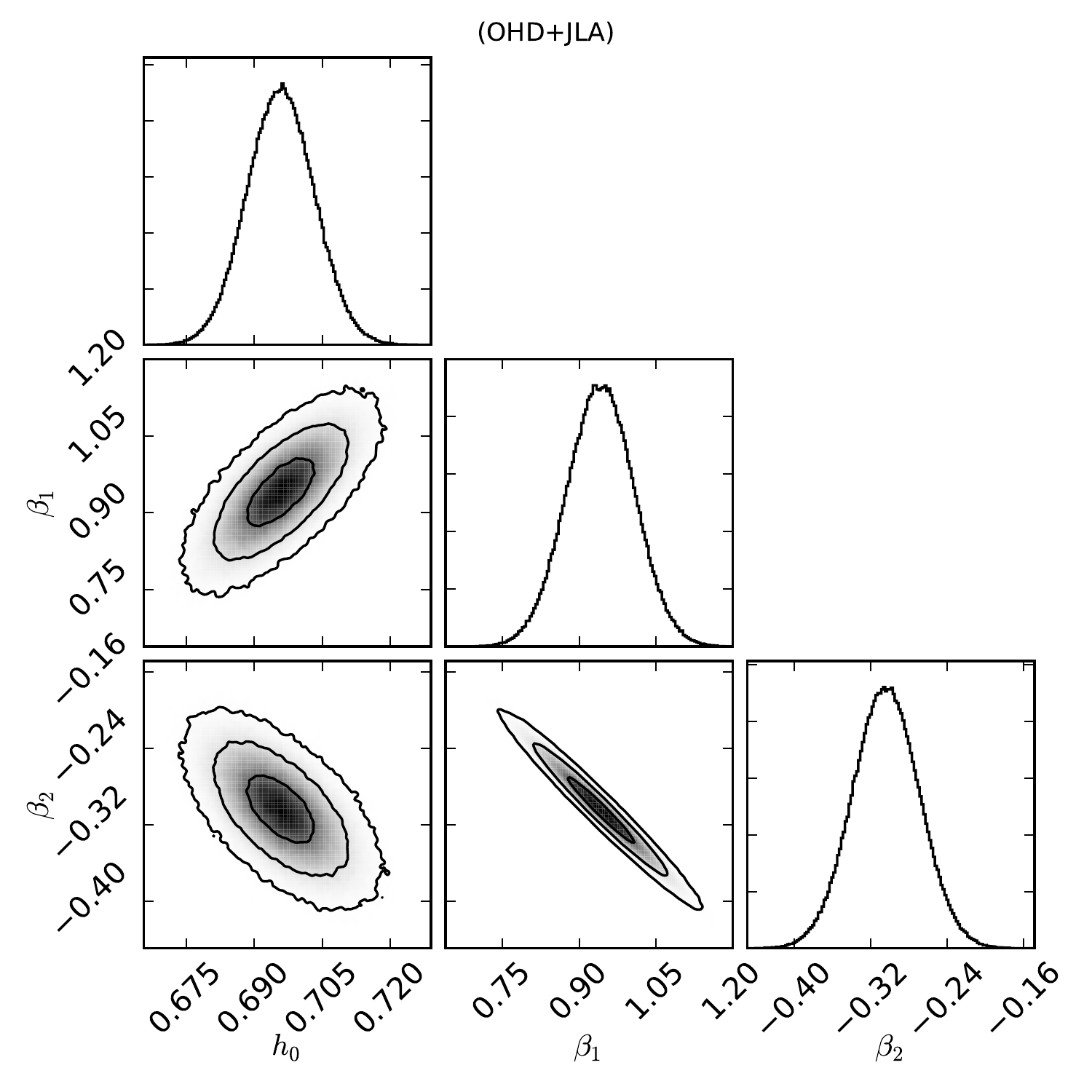}
\end{center}
\caption{{\small Confidence contours on the parameter space and the marginalized likelihood function of Model I, obtained from the combined analysis with OHD+JLA. The associated 1$\sigma$, 2$\sigma$, 3$\sigma$ confidence contours are shown.}}
\label{Model1contour}
\end{figure}
\begin{table}[tb]
\caption{{\small The parameter values and the associated 1$\sigma$ uncertainty of the parameters for the reconstructed model. The parameters are constrained in the combined analysis with JLA+OHD.}}
\begin{center}
\resizebox{0.7\textwidth}{!}{  
\begin{tabular}{c |c |c |c } 
  \hline
 \hline
   & $h_0$ & $\beta_1$ &  $\beta_2$ \\ 
 \hline
 Model I & $0.696^{+0.007}_{-0.007}$ & $0.942^{+0.065}_{-0.066}$ & $-0.304^{+0.035}_{-0.034}$\\ 
 \hline
 Model II & $0.700^{+0.008}_{-0.008}$  &  $0.737^{+0.042}_{-0.042}$ & $-0.906^{+0.102}_{-0.102}$\\ 
 \hline
 Model III & $0.700^{+0.008}_{-0.008}$ & $-0.170^{+0.064}_{-0.064}$ & $0.907^{+0.102}_{-0.103}$\\ 
 \hline
  \hline
\end{tabular}
}
\end{center}
\label{resulttable}
\end{table}

\begin{figure}[htb]
\begin{center}
\includegraphics[angle=0, width=0.7\textwidth]{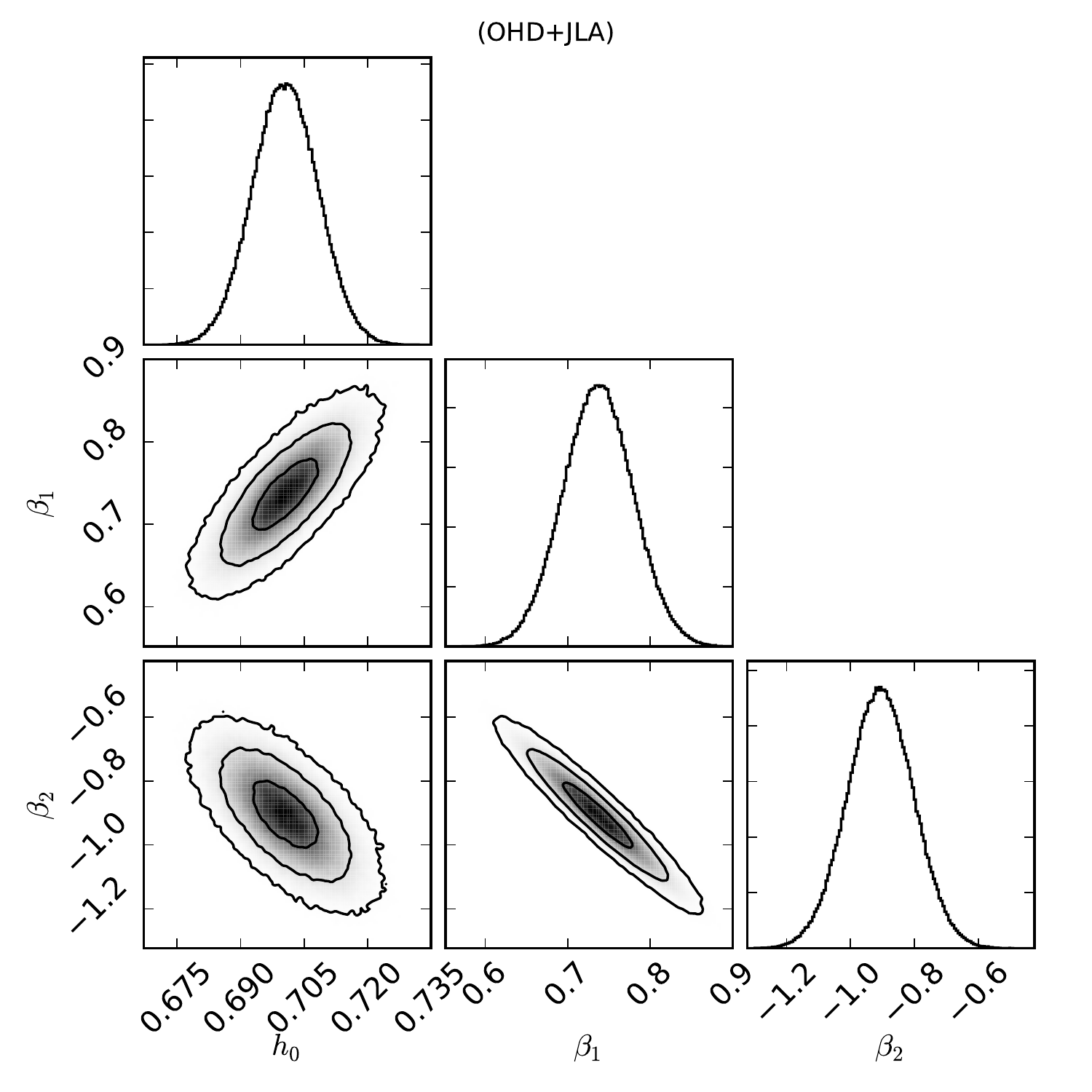}
\end{center}
\caption{{\small Confidence contours on the parameter space and the marginalized likelihood function of Model II, obtained from the combined analysis with OHD+JLA. The associated 1$\sigma$, 2$\sigma$, 3$\sigma$ confidence contours are shown.}}
\label{Model2contour}
\end{figure}

\begin{figure}[htb]
\begin{center}
\includegraphics[angle=0, width=0.7\textwidth]{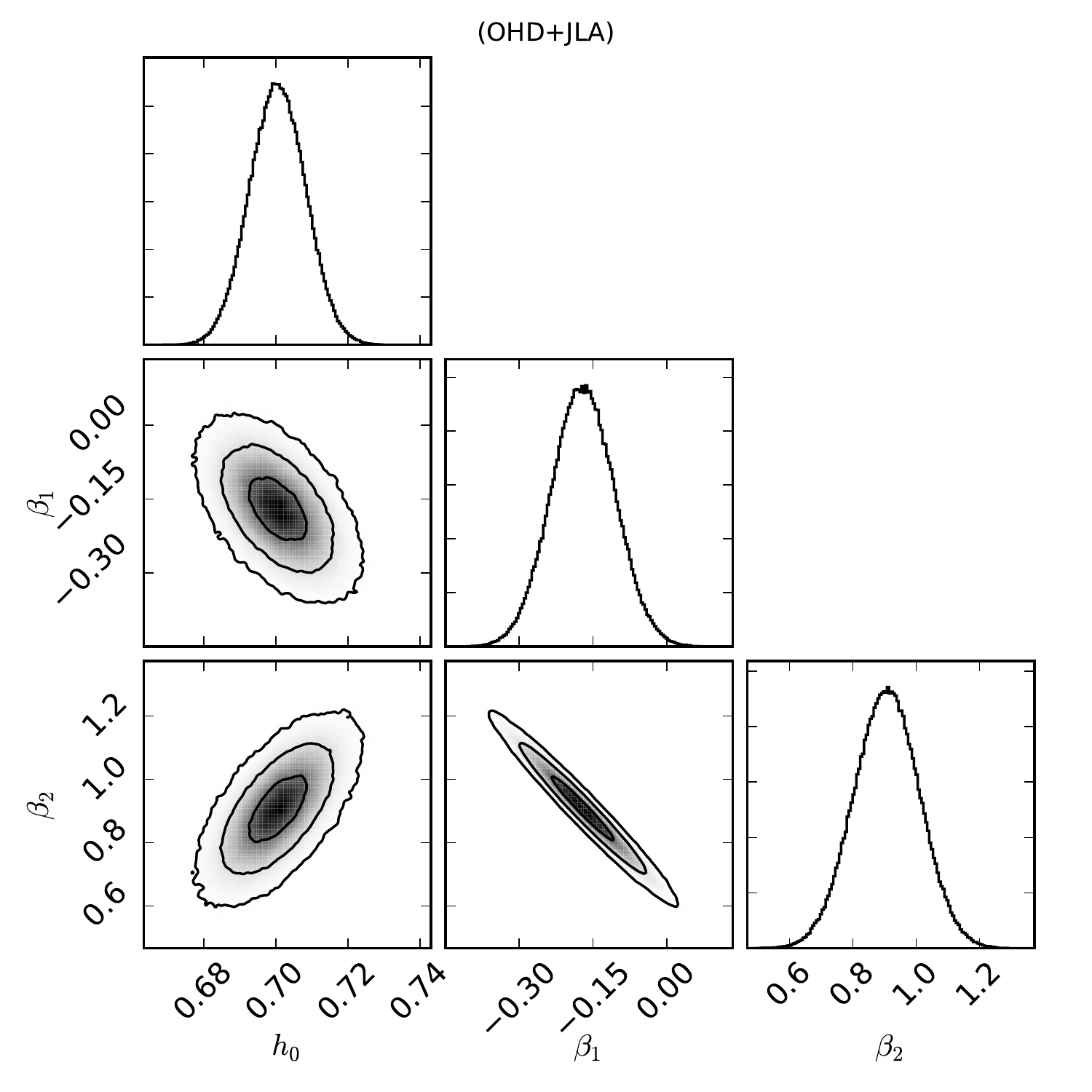}
\end{center}
\caption{{\small Confidence contours on the parameter space and the marginalized likelihood function of Model III, obtained from the combined analysis with OHD+JLA. The associated 1$\sigma$, 2$\sigma$, 3$\sigma$ confidence contours are shown.}}
\label{Model3contour}
\end{figure}

Figure \ref{Model1contour} shows the confidence contours on the parameter space and the marginalized likelihood function of Model I obtained in the combined analysis with OHD+JLA. The confidence contours and the likelihood functions of the parameters of Model II and Model III are presented in figure \ref{Model2contour} and \ref{Model3contour}. Table \ref{resulttable} shows that values of the parameters, obtained in the analysis with OHD+JLA for the reconstructed models. 

\par In the statistical analysis, the value of the parameters $\beta_1$ and $\beta_2$, which are actually the model parameters scaled by the present value of the coincident parameter $r$, are estimated. The value of $r$, estimated from the Planck measurement of $\Omega_{\Lambda}$, is $r=0.445\pm 0.010$. The coincident parameter $r$ remains constant in case of a holographic dark energy with Hubble scale cut-off in a spatially flat FLRW universe \cite{senpav}. Thus it can potentially resolve the {\it coincidence problem} of the standard model of cosmology.

\par Figure \ref{Model1contour} shows that in Model I, the Hubble parameter ($h_0$) has a positive correlation with the parameter $\beta_1$ and has a negative correlation with the parameter $\beta_2$. On the other hand, the parameters $\beta_1$ and $\beta_2$ are negatively correlated. Figure \ref{Model2contour} shows a similar correlations between the parameters for Model II, that is the $h_0$ has  positive correlation with $\beta_1$ and has negative correlation with $\beta_2$, similarly $\beta_1$ and $\beta_2$ are negatively correlated.  In Model III, $h_0$ has negative correlation with $\beta_1$ and positive correlation with $\beta_2$, and model parameters $\beta_1$ and $\beta_2$ are negatively correlated (figure \ref{Model3contour}).

From the conservation equation of dark matter density (equation (\ref{matcon})), one can see that the interaction term $Q$ can be absorbed in the left hand side of the equation. Thus we can assign a non-zero effective equation of state parameter for the dark matter also. Equation (\ref{matcon}) will look like 
\be
\dot{\rho}_m+3H\rho_m(1-\frac{\alpha}{r}) =0.
\label{wmeff1}
\ee 
The effective equation of state parameter of dark matter ($w_{eff}^m$) is thus given by $w_{eff}^m = -\frac{\alpha}{r}$.

With the estimated vales of $\beta_1$ and $\beta_2$, where $\beta_i = \frac{\alpha_i}{r}$ (from Table 1), one can estimate the values of $w_{eff}^m$ at $z=0$ as $-0.638, -0.737$ and $-0.737$ for the models I, II and III respectively. These values are far too less to have a dark matter contribution at present. In distant past, for $z$ close to a thousand, the effective equation of state parameter for the matter density $w_{eff}^m$ would have values close to $0.3$, $0.169$ and $0.170$ for Models I, II and III respectively. We shall come back to these at the end of the following section.


\section{Evolution of cosmological parameters}
\label{cospar}

\par
\begin{figure}[tb]
\begin{center}
\includegraphics[angle=0, width=0.32\textwidth]{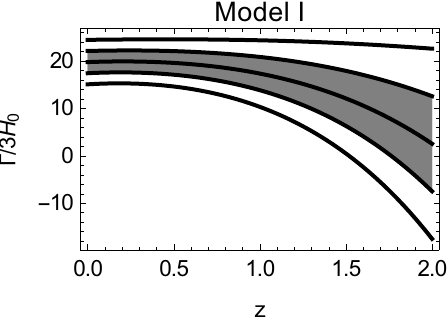}
\includegraphics[angle=0, width=0.32\textwidth]{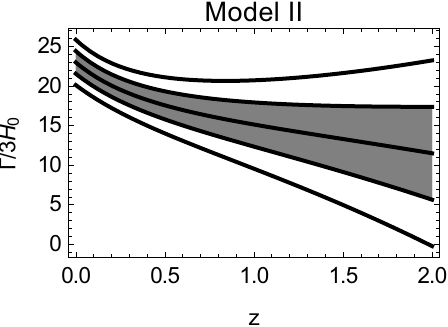}
\includegraphics[angle=0, width=0.32\textwidth]{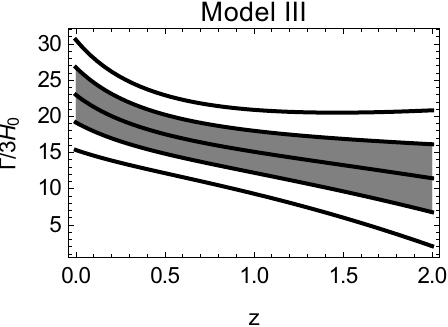}
\end{center}
\caption{{\small Plots of the interactions rate $\Gamma$ scaled by $3H_0$ for the reconstructed models. The best fit values and the associated 1$\sigma$, 2$\sigma$ confidence regions are obtained from the combined analysis with OHD+JLA.}}
\label{gamma}
\end{figure}

 The rate of interaction between dark energy and dark matter is defined as $\Gamma=Q/\rho_H$  \cite{senpav, am} and thus it can be expressed as,

\be
\Gamma=3H(z)\alpha(z).
\ee
 The rate of energy transfer and also the direction of energy flow depend on this term. Figure \ref{gamma} shows the plots of the interaction rate, $\Gamma$, scaled by 3$H_0$, for the reconstructed models. The plots show the interaction rate and consequently, the interaction function $Q$ remains positive. That means the energy gets transferred from dark energy to dark matter. It is consistent with the thermodynamic requirement discussed by Pavon and Wang \cite{pavonwang}. Plots of $\Gamma(z)$ show that it evolves in a very different way for Model I than that of Model II and Model III.

\begin{figure}[tb]
\begin{center}
\includegraphics[angle=0, width=0.32\textwidth]{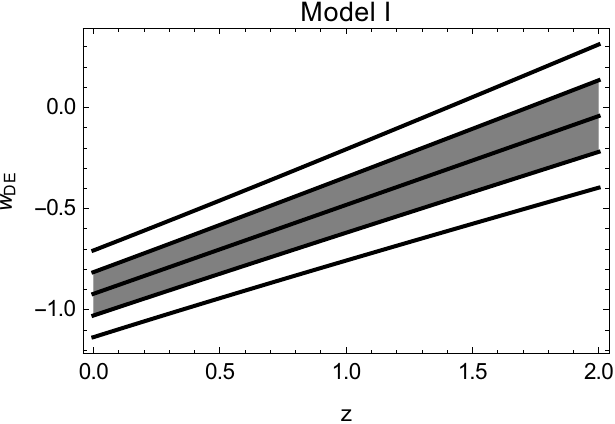}
\includegraphics[angle=0, width=0.32\textwidth]{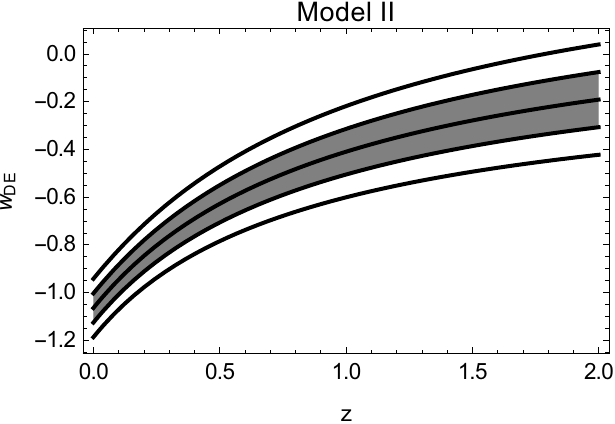}
\includegraphics[angle=0, width=0.32\textwidth]{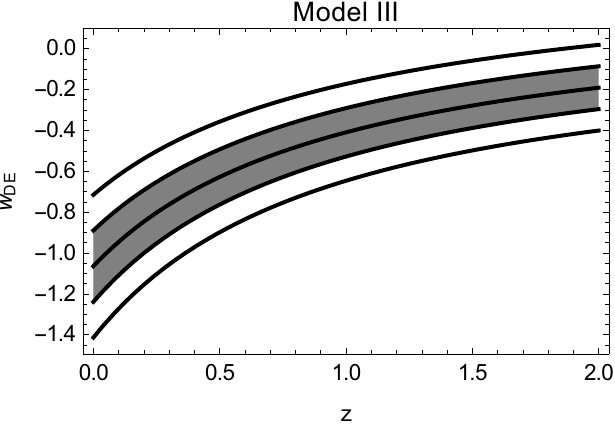}
\end{center}
\caption{{\small Plots of the dark energy equation of state for the reconstructed models. The best fit and 1$\sigma$, 2$\sigma$ confidence regions, obtained in the combined analysis with OHD+JLA, are shown.}}
\label{wdeplot}
\end{figure}

\begin{figure}[tb]
\begin{center}
\includegraphics[angle=0, width=0.32\textwidth]{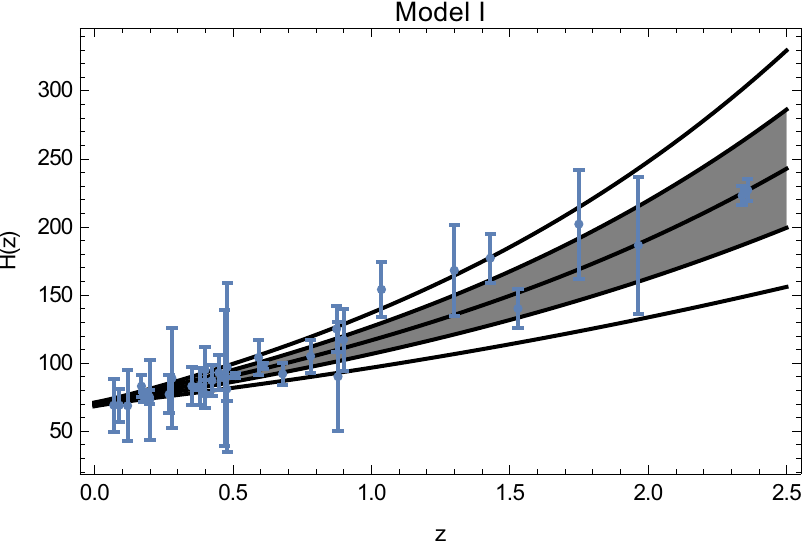}
\includegraphics[angle=0, width=0.32\textwidth]{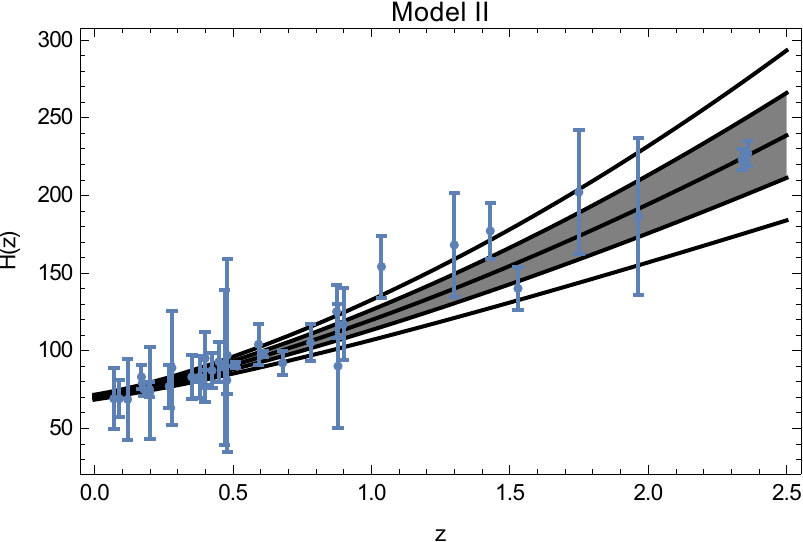}
\includegraphics[angle=0, width=0.32\textwidth]{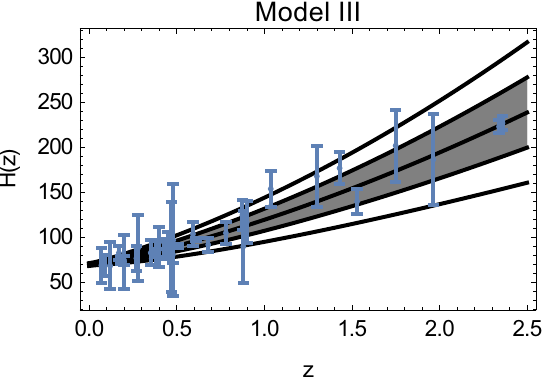}
\end{center}
\caption{{\small Plots of the Hubble parameter $H(z)$ for the reconstructed models and the observational data points along with the error bars. The best fit values and the associated 1$\sigma$, 2$\sigma$ confidence regions are obtained from the combined analysis with OHD+JLA.}}
\label{hubb}
\end{figure}
\begin{figure}[tb]
\begin{center}
\includegraphics[angle=0, width=0.32\textwidth]{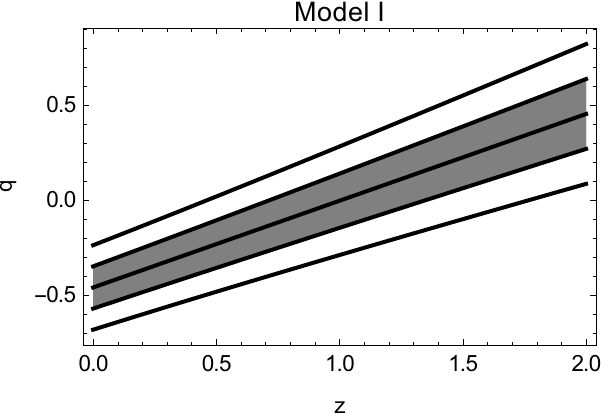}
\includegraphics[angle=0, width=0.32\textwidth]{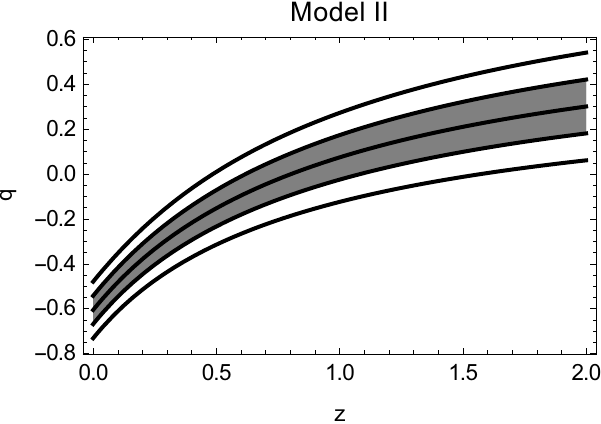}
\includegraphics[angle=0, width=0.32\textwidth]{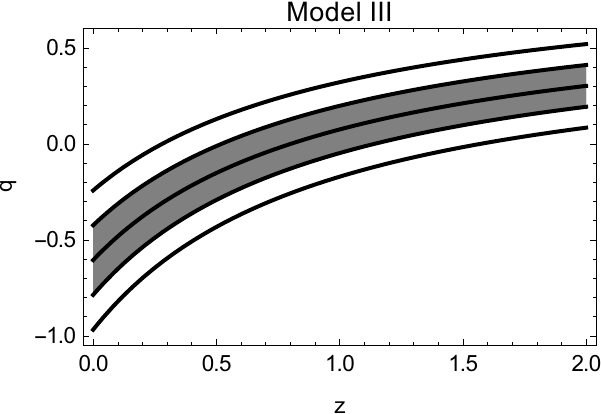}\\
\includegraphics[angle=0, width=0.32\textwidth]{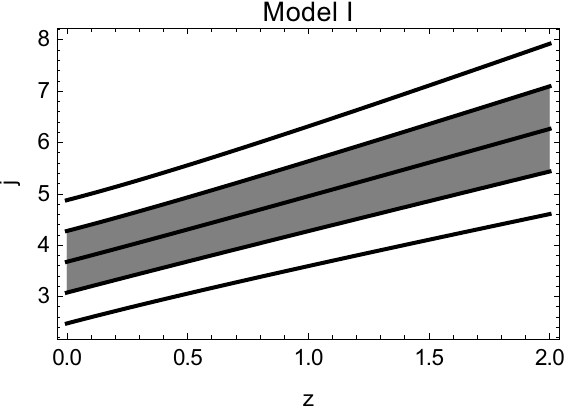}
\includegraphics[angle=0, width=0.32\textwidth]{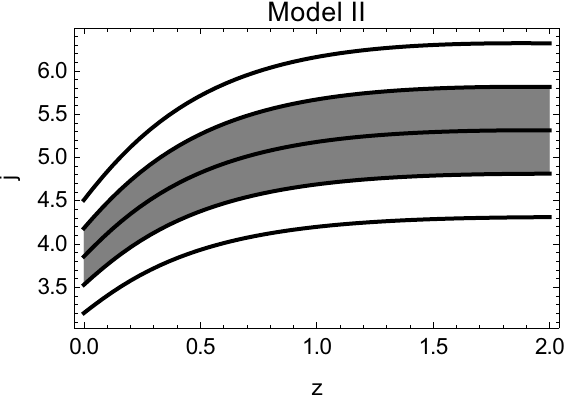}
\includegraphics[angle=0, width=0.32\textwidth]{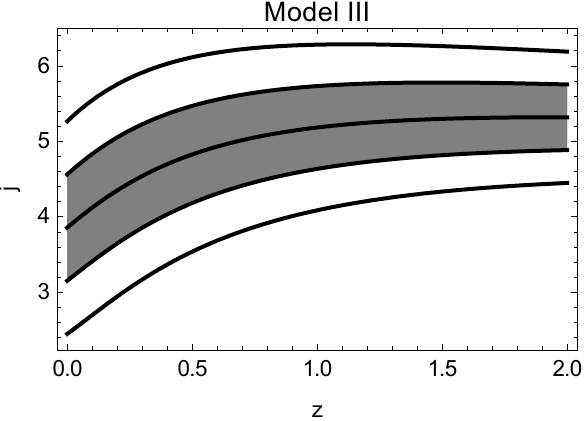}
\end{center}
\caption{{\small Plots of the deceleration parameter (upper panels) and the cosmological jerk parameter (lower panels) for the reconstructed models.}}
\label{qzplot}
\end{figure}

The dark energy equation of state parameter $w_{DE}$ is related to the total or effective equation of state parameter $w_{eff}=p_{total}/\rho_{total}$ as,
\be
w_{DE}=w_{eff}(1+r).
\ee
The $w_{DE}(z)$ evolves in the similar way to that of $w_{eff}(z)$ as $r$ is a constant for these models. Figure \ref{wdeplot} shows the evolution of the dark energy equation of state parameter for these models. The $w_{DE}(z)$ decreases at high redshift.

\par The evolution of the expansion rate, namely the $H(z)$, for the reconstructed models are shown in figure \ref{hubb} along with the observational data points. It shows that the $H(z)$ measurements from Lyman-$\alpha$ forest \cite{ohdLya} at redshift $z=2.34$ is well within the 1$\sigma$ confidence region of the reconstructed models. The deceleration parameter $q(z)$ plots (upper panels figure \ref{qzplot}) shows that it increases with redshift and there is a transition in the signature of $q(z)$. For Model II and Model III, the transition redshift $z_t<1$ which is consistent with direct observational result \cite{riess2004,fqrat}, but Model I shows the transition at a much higher redshift. Thus, Model I is not consistent with the observed evolution of $q(z)$ and it can be ruled out. As the deceleration parameter, which is the second order time derivative of the scale factor, is now an observable quantity and the evolution is highly degenerate for viable dark energy models, it is important to investigate the next order derivative of the scale factor to distinguish between the models. The third order time derivative of the scale factor, represented in a dimensionless way, is called the jerk parameter, defined as,

\be
j=\frac{1}{aH^2}\frac{d^3a}{dt^3}.
\ee
It remains constant for $\Lambda$CDM model and the value is unity. For the models, for which the $\Lambda$CDM can be recovered as a limiting case, the value of jerk parameter hovers around the corresponding $\Lambda$CDM value \cite{ammnras}. The models, reconstructed in the present work, show a highly different evolution of the jerk parameter (lower panels of figure \ref{qzplot}). 

Thus the cosmographical parameters like $q$ etc are indeed consistent with the observations. But the problem lies with the physical quantities like $w_{eff}^m$. In the last section we saw that the present value of $w_{eff}^m$ is negative in all the the three models. For a high redshift,  $w_{eff}^m$ is positive, but in the case of Model I, this is too high to describe any reasonable matter sector. For Models II and III, $w_{eff}^m$ picks up values $.169$ and $.170$ respectively. This appears to be reasonable , but high enough to produce a substantial pressure and the matter cannot be cold. It is intriguing to note from figure \ref{wdeplot} that $w_{DE}$ allows a zero value in at least a 1$\sigma$ or at least a 2$\sigma$ level, as if that can, within the error bar, give rise to a cold dark matter! However, that does not resolve the issue in any way as in that kind of an identification, the present behaviours of both the matter sector will be like a dark energy.

\section{Bayesian analysis for model selection}
\label{bayev}

In a Bayesian analysis, statistical preference of a model is judged by calculation the Bayesian evidence. It is defined as,

\be
E=\int (Prior\times Likelihood)d\theta_1d\theta_2.........d\theta_n,
\ee

where $\theta_i$'s are the model parameters. A higher value of the evidence indicates the preference toward a model compared to others. One model can be judged as superior to the other according to Bayesian analysis if the value of the first one is around $10^3$ order of magnitude higher than the other one. In the present case, all the three models have three parameters, namely $\beta_1$ and $\beta_2$, and the present Hubble parameter $h_0$. A uniform prior has been assumed for the model parameters. The posterior probability distribution of the parameter is proportional to the likelihood function if the prior probability is uniformly distributed over the parameter range. The evidence values, calculated for the reconstructed models, are

\be
Model~~I.~~~~~~~E_1=P_1\int Likelihood.d\beta_1d\beta_2dh_0=9.823\times 10^{-22},
\ee 

\be
Model~~II.~~~~~~~E_2=P_2\int Likelihood.d\beta_1d\beta_2dh_0=4.123\times 10^{-13},
\ee 

\be
Model~~III.~~~~~~~E_3=P_3\int Likelihood.d\beta_1d\beta_2dh_0=4.726\times 10^{-13},
\ee

where $P_1$, $P_2$ and $P_3$ are the prior probability distribution for the respective models. The Bayesian evidence values clearly show that Model I is actually ruled out compared to the other two. On the other hand, Model II and Model III are very close according to statistical model selection. This result is consistent with our conclusion about the models from the evolution of the Hubble parameter and deceleration parameter. We have also calculated the evidence for spatially flat $\Lambda$CDM model using the same sets of data and obtained $E_{\Lambda CDM}=1.475\times 10^{-15}$. This is definitely less than the evidence values of the reconstructed Model II and Model III ! It deserves mention that we have included the $H(z)$ measurements from Lymann-$\alpha$ forest which is known to have some discrepancy with the $\Lambda$CDM cosmology \cite{Ade:2015xua}. This may be the reason for the better performance of the two interacting models than the $\Lambda$CDM against reliability.

\section{Conclusion}
\label{dis}

In the present work, a holographic dark energy model, with the Hubble radius being the IR cut-off, has been reconstructed for three different choices of a parameter that determines the strength of the interaction between the dark matter and the dark energy. It has already been discussed that if the interaction function is assumed to be proportional to $H\rho_H$, then the coupling parameter ($\alpha$) needs to vary with time for a successful transition from the decelerated to the accelerated phase of expansion. 

\par The nature of the interaction rate in Model I is qualitatively different from that of Model II and Model III.  The interaction rate increases significantly at lower redshift ($z<0.5$) for both Model II and Model III (figure \ref{gamma}). On the other hand, for Model I, it decreases with the evolution.

\par In the present analysis, the viability of the interaction is obtained. Results show that $\alpha\approx(1+z)$ type of models (Model I) are not consistent with the observed evolution scenario. On the other hand, Model II and Model III are highly consistent with the observed nature of the deceleration parameter (upper panels \ref{qzplot}). A Bayesian analysis for model selection also shows quite conclusively that Model I is ruled out in comparison with Model II and Model III. The latter two, in fact, do better than a $\Lambda$CDM model.

\par This indicates that for viable interacting holographic dark energy models, $\alpha$ has an inverse dependence on $z$, the interaction is smaller in the past, i.e., at higher redshift.

\par The magnitude of the dark energy equation of state parameter ($w_{DE}$) is small at high redshift for these models (figure \ref{wdeplot}) and even allows a zero value within 2$\sigma$ . It eventually decreases to values negative enough to generate the accelerated expansion. The present value of $w_{DE}$ remains in non-phantom regime ($w_{DE}>-1$) for Model I. For Model II and Model III, however, it is in the phantom regime ($w_{DE}<-1$). The interaction rate ($\Gamma$) (figure \ref{gamma}) and consequently the interaction function $Q$ remains positive which indicates that the pumping of energy is from the dark energy component to the dark matter component. Pavon and Wang \cite{pavonwang} have shown that $Q$ should be positive as a thermodynamic requirement.

\par The present value of the cosmological jerk parameter obtained for these models remains between $3$ to $5$ at 1$\sigma$ level which shows a strong departure from the corresponding $\Lambda$CDM value of unity. Thus the interacting holographic dark energy models, reconstructed in the present work, are indeed distinguishable from the models close to the $\Lambda$CDM for which the jerk parameter value hovers around unity.

\par The values of the model parameters are estimated in a Markov Chain Monte Carlo (MCMC) analysis with various observational datasets. The cosmic microwave background (CMB) shift parameter measurement has not been used. Actually, the shift parameter is estimated with the fiducial assumption of $\Lambda$CDM cosmology and hence can introduce bias in the analysis.

\par The parameters $\beta_1$ and $\beta_2$ are strongly negatively correlated and thus contribute to $\alpha$ in the opposite sense. However, their values are not close for Models I and III (see figures 1 and 3) so they do not nullify each other for small values of $z$. For Model II, $\beta_1$ and $\beta_2$ are indeed close, but near $z=0$, the term containing $\beta_2$ does not contribute.

\par Although the three models chosen for the interaction function are not at all general, the results obtained in this work very clearly indicate that the interaction in the dark sector is sizable at the present epoch rather than in the past, as Model I is ruled out both by the observational data directly and also by the Bayesian evidence analysis. 

\par Although Models II and III very efficiently describe the kinemtatical parameters, the actual composition of the universe is hardly consistent with the models. It gives a negative value for  $w_{eff}^m$ as if the universe at present is composed only of a dark energy. The composition of the early universe also is not properly described by these models. The problem arises because for the holographic dark energy models with Hubble horizon as the IR cut-off in a spatially flat universe, the { \it coincidence parameter} is essentially a constant \cite{senpav}. Thus we clearly conclude that holographic dark energy models with Hubble scale IR cut-off do not allow a spatially flat universe. The spatial geometry of the universe must have some curvature in this case.

\vskip 1.50 cm

{\bf Acknowledgment}\\
The authors would like to thank the anonymous referee, whose suggestions led to a qualitative improvement of the paper.

\end{document}